# Growth and hole density control through equilibrium oxygen annealing of optimally doped $Y_{1-x}Ca_xBa_2Cu_3O_{7-\delta}$ single crystals

Zheng Wu[*] and Pei-Herng Hor[+]
Department of Physics and Texas Center for Superconductivity at the University of Houston, Texas 77204-5002

*email: zwu@uh.edu
+email: phor@uh.edu

**Abstract.** We have grown calcium and oxygen co-doped $Y_{1-x}Ca_xBa_2Cu_3O_{7-\delta}$ (CD-Y123) single crystals using self-flux method. A method of fine tuning oxygen content to reach the equilibrium state through *in situ* monitoring of the conductivity change is established. Structural, compositional and electronic property characterizations of optimally co-doped $Y_{1-x}Ca_xBa_2Cu_3O_{7-\delta}$ crystals indicate that they are high quality equilibrium crystals with a sharp superconducting transition width of 0.2K.

## 1. Introduction

High temperature superconductivity is induced by doping holes into $CuO_2$ planes using either cation dopants or anion dopants. $La_{2-x}Sr_xCuO_{4+\delta}$, for instance, can be doped by either cation dopant strontium ($Sr^{++}$) or anion dopant oxygen ($O^-$). Dopand Sr, replaces La, occupies the lattice site such that it behaves as a quenched disorder. Dopant O enters into interstitial sites and is mobile down to ~200 K so that it behaviors as an annealed disorder. In Sr and O co-doped $La_{2-x}Sr_xCuO_{4+\delta}$ (CD-La214) different disordering effects, quenched disorder vs. annealed disorder, can be tuned at a fixed hole density by tuning Sr content vs. O content. A true intrinsic physical property will reveal itself as a property that depends only on planar doped-hole density $P_{pl}$ defined as the holes per Cu atom on the $CuO_2$ planes; it should be independent of the nature, cation or anion, and the ordering of the dopants. Systematic studies of the preparations and oxygen doping efficiency of CD-La214 polycrystalline samples indicated that delicate electrochemical oxidation performed at elevated temperature and long time post annealing are required in order to achieve thermodynamically equilibrium samples up to the optimal doping concentration[1,2]. Equilibrium polycrystalline samples have only two superconducting transitions with $T_c$ = 15 K and $T_c$ = 30 K [2] which is independently confirmed later [3]. Systematic high pressure studies, when pressure was applied at room temperature, of pure oxygen doped $La_2CuO_{4+\delta}$ (OD-La214) and CD-La214 indicate that they share a common phase diagram: a single $T_c \approx$ 30 K transition regime with an unusually large pressure coefficient $dT_c^{(30)}/dp \approx$ 10 K/GPa for $P_{pl}$ < 0.085 and a double, $T_c \approx$ 30 K followed by a $T_c$ = 15 K, transitions regime with $dT_c^{(15)}/dp \approx$ -4 K/GPa for $P_{pl}$ > 0.085. Both the phase boundary and the pressure coefficients in different regimes were independent of the contents of anion and cation dopants, they depended only on the $P_{pl}$ [4]. These two intrinsic superconducting transitions are manifestations of two intrinsic electronic superconducting phases that are independent of the nature and the ordering of the dopants.  High pressure Hall effect measurements of OD-La214 at room temperature exhibit consistent hole localization at $P_{pl}$ = 0.06 (~1/16) and $P_{pl}$ = 0.100 (~1/9) under various pressures [5]. Indeed, far infrared charge dynamic studies of CD-La214 suggested that formation of charged ordered states at $P_{pl}$ = 1/16 and $P_{pl}$ = 2/16 with $T_c$ = 15 K and $T_c$ = 30 K, respectively [6]. Furthermore in a series of specifically prepared high quality pure Sr doped $La_{2-x}Sr_xCuO_4$ (SD-La214) single crystals we observed a much sharper, see figure 2 in reference [7], $T_c$ = 15 K at $P_{pl}$ = 0.063 and $T_c$ = 30 K at $P_{pl}$ = 0.11 transition width ($\Delta T$ ~ 2 K) than that ($\Delta T$ ~ 6 K) for transition temperature other than these two intrinsic superconducting transitions. Charge dynamic studies of these SD-La214 single crystals were intrinsically consistent with the picture obtained from the polycrystalline CD-La214 in reference [8]. All these results strongly suggest that there are intrinsic superconducting phases at magic doping concentrations $P_{pl}$ = $m/n^2$, m and n are positive integers and n =3 or 4 with m ≤ n.  Detailed studies of these intrinsic electronic superconducting phase shall shed lights on the mechanism of high $T_c$ cuprates. Unfortunately preparation of equilibrium high quality CD-La214 crystals turns out to be, if not impossible, extremely difficult; micro-cracks developed upon oxygen charging and the crystal can even disintegrate at high oxygen doping [9]. This shortfall had seriously hindered the detailed studies of the intrinsic superconducting phases. The crystal is not in an electronic equilibrium state and results derived from CD-La214 crystals with $T_c$ different from these two intrinsic transitions should be interpreted with care.

A parallel situation occurred in the $Y_{1-x}Ca_xBa_2Cu_3O_{7-\delta}$ high $T_c$ superconductors. For pure oxygen doped $YBa_2Cu_3O_{7-\delta}$ (OD-Y123) the doped hole density of in $CuO_2$ plane can be tuned from almost zero doping up to slightly overdoped regime by adjusting the oxygen content and the chain ordering [10,11]. Two intrinsic transitions were identified in the OD-Y123 system at $T_c$ = 60 K or $T_c$ = 90 K and the $T_c$ versus doping behavior depended only on the hole density [12]. Calcium (Ca) and oxygen (O) co-doped $Y_{1-x}Ca_xBa_2Cu_3O_{7-\delta}$ (CD-Y123) can extend the doping well into the overdoped regime with reduced orthorhombic strain due to oxygen ordering, at a price of increasing out-of-plane disorder due to Ca dopant. Different disordering effects, quenched disorder vs. annealed disorder, can be tuned at a fixed hole density by tuning Ca content vs. O content. Therefore, preparing high quality equilibrium CD-Y123 single crystals and fine-tuning their oxygen content provides a unique opportunity in studying the intrinsic electronic properties of high $T_c$ cuprates.

In this paper we present the growth and characterizations of CuO-BaO self-flux grown optimal CD-Y123 crystals. All of the full-width-at-half-maximum (FWHM) of x-ray rocking curves (006 peak) of our crystals are less than 0.05º. Based on an oxygen annealing process we established, sharp transitions with width of 0.2 K, defined as the temperature interval for a drop of normal state value from 10% to 90%, were observed by thermoelectric power (TEP) and resistivity measurements. The magnetic measurements show that the shielding (zero-field cooled) signal is about 100% and the Meissner (field-cooled) signal is as large as 45% of that of a perfect superconductor. Those are best results for optimal CD-Y123 crystals can be found in the literature so far.

## 2. Experimental Details

The $Y_{1-x}Ca_xBa_2Cu_3O_{7-\delta}$ single crystals were grown using a self-flux method, which have been proven to be the best method for growing OD-Y123 single crystals. $BaZrO_3$ crucibles were used since it does not react with the melts and consistently yields high quality OD-Y123 crystals [13,14].

High purity $Y_2O_3$ (99.999), $BaCO_3$ (99.997%), CuO (99.995%), and $CaCO_3$ (99.995%) were used as starting materials. We fix the composition of the flux at the eutectic point of BaO and CuO with molar ratio BaO:CuO = 28:72. The starting compositions to grow $Y_{1-x}Ca_xBa_2Cu_3O_{7-\delta}$ single crystals are 10~12% [$1/6(Y_{1-x}Ca_xBa_2Cu_3O_7)$] and 90~88% [$(BaO)_{0.28}(CuO)_{0.72}$]

A mixture of the starting composition was first ball milled and calcined at 850°C for 10 hours followed by another thorough ball milling before packing into the crucible. About 50g of mixed powder was pressed into a ~19 ml $BaZrO_3$ crucible using a pestle. The crucible was placed in a muffle furnace with heating elements mounted on four side walls. A temperature gradient $\Delta T$ = 6-8 K over the crucible ($\phi$ = 30 mm) was established by adjusting the distance between the crucible and the back wall of the furnace. The temperature gradient was monitored by Pt/Pt-10Rh thermocouple configured such that the temperature difference is measured directly to reduce random noise and improve signal-to-noise ratio. The typical temperature fluctuation observed is ~ ±0.1°C. The molten mixture was hold at 1040°C for 16 hours then fast cooled at a rate of 4°C per minute to 1000°C. After $\Delta T$ was established, the furnace is cooled at a rate of 0.5~2.3°C per hour from 1000°C to 945°C where the flux was decanted inside the furnace.

The crystal perfection was examined by rocking curves using a Rigaku Geigerflex X-Ray Powder Diffractometer equipped with a curved graphite single crystal monochromator. Chemical compositions of the crystals were analyzed by wavelength dispersive spectrometry (WDS) using a JEOL JXA8600. An ac thermoelectric power measurement method [15] was employed to determine planar hole density. The anisotropic resistivity of crystals was measured based on an 8-lead method developed by Levin [16]. DC magnetization data was measured using a Quantum Design Magnetic Properties Measurement System (MPMS).

3. Results and Discussions

All of the crystals were characterized by measuring the rocking curves of the (006) peak before further studies. In figure 1(a) we show a crystal's rocking curve of (006) peak with FWHM=0.046°. One can expect smaller FWHM if the rocking curve was taken by a system equipped with a double-crystal monochromator.

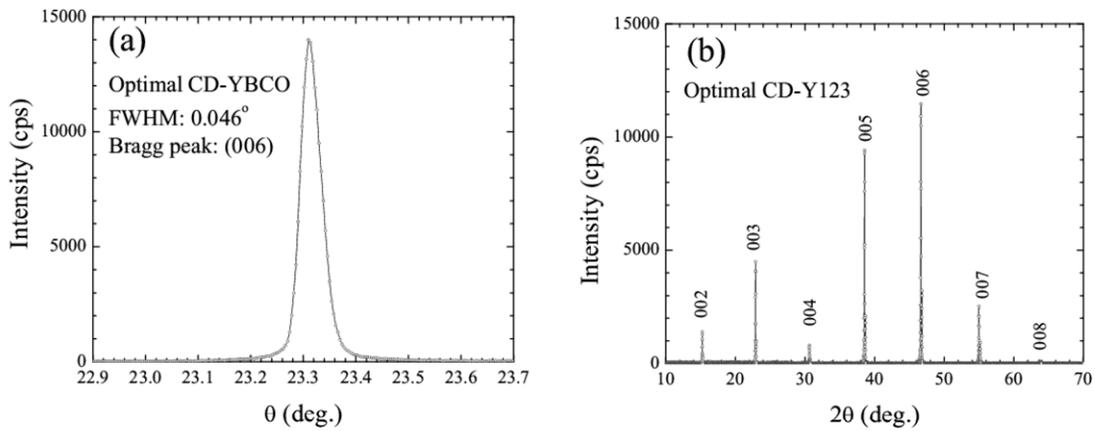

Figure 1. (a) Rocking curve of (006) reflection and (b) $\theta$-$2\theta$ scan for optimal CD-Y123 crystal with $x = 0.10$. The surface dimension of the crystal is 1.16 mm × 0.20 mm and the x-ray beam covered the entire surface.

The phase purity was checked by x-ray $\theta$-$2\theta$ scan. The diffraction pattern in figure 1(b) shows that all the diffraction peaks can be indexed with ($00l$) crystal planes of Y123 indicating it is a single grain with no impurity phase inclusions. Composition analysis using high-precision electron-probe microanalysis (EPMSA) with WDS showed that cation stoichiometry of

(YCa):Ba:Cu was 1:2:3 within the experimental accuracy of ±2%. The crystals have a calcium content ranging from $x = 0.07$ to $x = 0.18$. The variation of Ca content from point to point on one crystal is less than ±1.8%.

In CD-Y123 system $P_{pl}$ depends on both Ca content ($x$) and O content (7-$\delta$). Although, without ambiguity, there are $x/2$ amount of holes coming from Ca doping, the determination of holes contributed by oxygen dopant is non-trivial. The doping efficiency of oxygen depends on Ca level and the relation $P_{pl}(x, \delta)$ is not well-established. We have proposed a universal hole scale based on the TEP value at 290 K ($S^{290 K}$) which determines the $P_{pl}$ in the CuO$_2$ planes [17]. $P_{pl}$ is a true measure of planar hole density that works for all cuprate superconductors [18]. All the $P_{pl}$ values reported in this paper were determined based on our universal hole sacle [17].

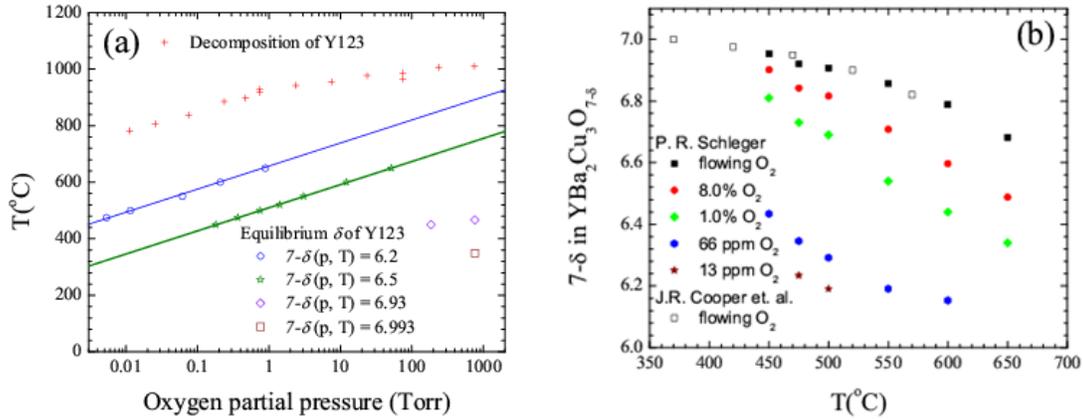

Figure 2. (a) Equilibrium temperature for oxygen content (7-$\delta$) and the decomposition temperature of OD-Y123 vs. oxygen partial pressures. (b) Oxygen content (7-$\delta$) in OD-Y123 vs. temperature at which samples are equilibrated for various oxygen partial pressures.

For a crystal with a specific $x$, its $P_{pl}$ can be tuned from $x/2$ to overdoped regime through adjusting oxygen deficiency $\delta$ by annealing crystal under specific oxygen partial pressure ($p_{O2}$) and temperature ($T$). The $\delta$ can be adjusted by fixing either one of the parameters and changing the other. In figure 2(a) we summarized various equilibrium temperatures of OD-Y123 reported in the literature into the $p_{O2}$-$T$-$\delta$ phase diagram. Open symbol is equilibrium temperature and oxygen partial pressure at several specific oxygen contents [19]. Among them, open stars ($\delta = 0.5$) are where *O-T* transition taking place [20,21]. The annealing temperature at each $p_{O2}$ is limited by decomposition temperature [22] which is also plotted (cross) in the phase diagram. We adjusted $\delta$ by fixing the $p_{O2}$ and change the annealing temperature. To achieve a certain *y*, we fixed $p_{O2}$ at 1 atm, 8.0% $O_2$ - 92.0%Ar, 1.0% $O_2$ – 99.0%Ar, and pure Ar (with $O_2$ < 0.3ppm) gases for different target regions. In figure 2(b), we plot *T* versus $\delta$ for OD-Y123 at various $p_{O2}$ collected from the literatures [19,23]. These two figures serves as a starting point for us to design annealing process to adjust $\delta$ in CD-Y123 crystals since both oxygen content [24] and oxygen doping efficiency [25] change with calcium concentration when annealing at fixed *T* and $p_{O2}$.

To ensure an equilibrium oxygen distribution we monitor the annealing process by *in situ* resistance measurement. Oxygen diffusion is mainly along the *ab*-plane direction which is about 5 to 6 orders of magnitude faster than that in the c-axis direction [26,27]. Therefore, we only need to concern *ab*-plane diffusion. We assume the initial oxygen concentration ($C_i$) of the crystal is uniform and all surfaces are kept at a constant concentration ($C_f$). The oxygen concentration at any time instant t, $C(t) = C$ is [28]

$$\frac{C - C_i}{C_f - C_i} = 1 - \frac{4}{\pi} \sum_{n=0}^{\infty} \frac{(-1)^n}{2n+1} \cos\left[\frac{2n+1}{2l}\pi x\right] e^{-\left(\frac{2n+1}{2l}\pi\right)^2 Dt} \quad (1)$$

where the sample thickness, along diffusion direction (*x*-axis), is 2*l* with the center at *x* = 0 and *D* is the diffusion coefficient.

The electrical conductivity for mobile hole carriers can be written as

$$\sigma = p_{pl}\mu e$$

Where $\mu$ is the hole mobility and *e* is the electron charge. At fixed *x* we assume the $P_{pl}$ changes linearly with oxygen concentration and $\mu$ keeps constant in the studied range of oxygen concentration during isothermal annealing; we have

$$\frac{\sigma(x,t)-\sigma_i}{\sigma_f-\sigma_i} = \frac{C(x,t)-C_i}{C_f-C_i} \quad (2)$$

Assuming $l$ is much shorter than the distance from the sample edge to the nearby voltage contact so that the diffusion along current direction will not contribute to the resistance between voltage contacts. Under this condition, measured resistance can be treated as parallel connection of infinitesimal $dR(x,t)$. Thus, the measured conductance ($G$) is the integration of $dG(x,t)$, which is proportional to $\sigma(x,t)dx$, over sample dimension along diffusion direction $x$ (width of the crystal). Therefore, the conductance varies with time as

$$\frac{G(t)-G_i}{G_f-G_i} = 1 - 8\sum_{n=0}^{\infty}\left[\frac{1}{(2n+1)\pi}\right]^2 e^{-(2n+1)^2 t/\tau} \quad (3)$$

where the diffusion time (relaxation time) constant $\tau = \frac{(2l)^2}{\pi^2 D}$ is determined by geometric parameter $l$ and diffusion coefficient $D$.

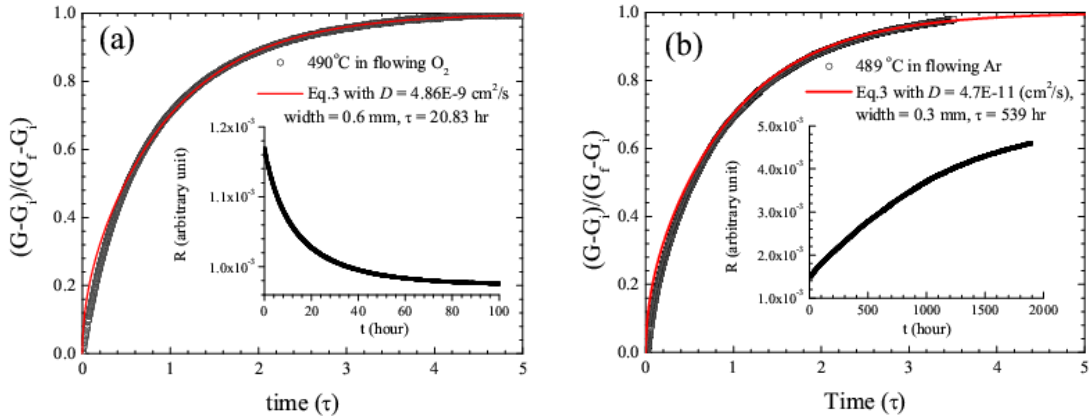

Figure 3. Normalized conductance vs. time for in-diffusion (in flowing oxygen) and out-diffusion (in flowing argon). Solid lines are simulations using equation (3). The insets are time dependence of resistance monitored during oxygen in- (a) and (b) out-diffusion.

In figure 3(a) and 3(b) we present examples of normalized conductance change versus time in units of time constant $\tau$ for crystals annealed in flowing oxygen and flowing argon environments, respectively. The red line is the fit to equation (3) with a diffusion coefficient of $D = 4.86\times10^{-9}$

cm$^2$/K in flowing oxygen at 490$^o$C and $D$ = 4.7×10$^{-11}$ cm$^2$/K in flowing argon at 489$^o$C. The oxygen out-diffusion in flowing Ar is two orders of magnitude slower than that of in-diffusion in flowing oxygen. The insets in figure 4 are the plots of resistance versus time during annealing process. Note that the resistance in flowing argon (inset of figure 4(b)) appears to be changing linearly with time during the initial several hundred hours, the linear change in resistance was also observed in OD-Y123 single crystals [29, 30]. It was attributed to a surface reaction limited out-diffusion process [31]. However in the long time annealing, 2000 hours in our case, we find that the overall resistance followed bulk diffusion limited process if we plot the result as normalized conductance change versus time (figure 4(b)). Equation (3) describes the experimental data for both in- and out-diffusion very well and can be used to judge the oxygen homogeneity in the crystal. In an annealed crystal the oxygen concentration at the center is 99.5% of that at the surface when the annealing time is five times of $\tau$ (93.7% for 3$\tau$). We conclude that it is necessary to express monitoring process by plot normalized conductance change versus diffusion time constant $\tau$. Thus, the oxygen distribution is only related to $\tau$ regardless the differences in sample dimension and diffusion coefficient. Otherwise, without knowing the dimensions of the sample, it is difficult to assess the annealing process and the homogeneity of the final state. We monitor the annealing process *in situ* and quench the crystal to room temperature when the 5$\tau$ criterion is satisfied.

An optimal CD-Y123 with $x$ = 0.10 (rocking curve and XRD presented in figure 1 (a) and 1(b)) obtained by annealing the crystal in flowing oxygen at several temperatures high than 500$^o$C step by step. In each step, the annealing time was long enough to ensure oxygen equilibrium state reached. Figure 4(a) shows the temperature dependence of TEP. The hole density is $P_{pl}$ = 0.256 which is very close to optimal doping concentration $P_{pl}$ = 0.25. The anisotropy resistivity is shown in figures 4(b) and 4(c). Both TEP and resistivity (in-plane and out-of-plane) exhibit sharp transitions with transition width, defined as 10% to 90% drop from normal state value, ~ 0.2 K (insets of figures 4(a), 4(b) and 4(c)). Temperature dependence of in-plane resistivity is linear from room temperature down to $T_c$, similar to that of the optimal OD-Y123. Resistivity value is smaller than but close to that along $a$ direction and large than that along $b$ direction (oxygen chain direction) of the optimal OD-Y123 [32, 33]. Out-of-plane resistivity was much larger (1.6 times at room temperature) than that of optimal OD-Y123 [33] which can be attributed to the increase of $c$-axis lattice constant due to the increasing of both $x$ [34] and $\delta$ [21].

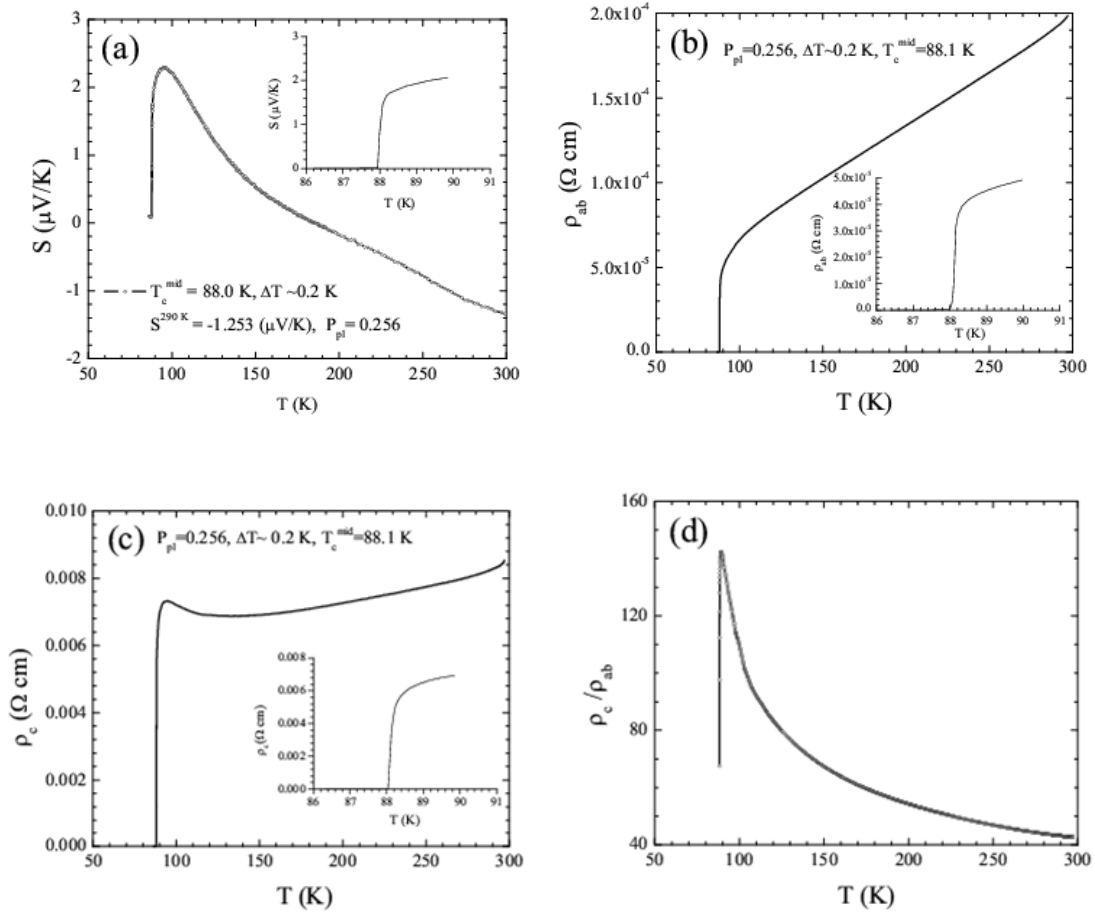

Figure 4. TEP (a), in-plane (b), out-of-plane (c) resistivity, and anisotropy vs. temperature. Insets of (a), (b), and (c) show sharp transition widths of 0.2 K for TEP, in-plane and out-of-plane resistivities.

We also performed the magnetization measurements on an optimally doped crystal ($P_{pl}$ = 0.247) with $x$ = 0.11. We obtained 0.3 K and 1 K transition widths (same definition as that used for TEP and resistivity) for shielding (zero-field cooled) and Meissner (field-cooled) signals, respectively. The absolutely susceptibility of shielding signal was about 100% and Meissner signal was as large as 45% of that of a perfect superconductor (figure 5) after demagnetization factor correction.

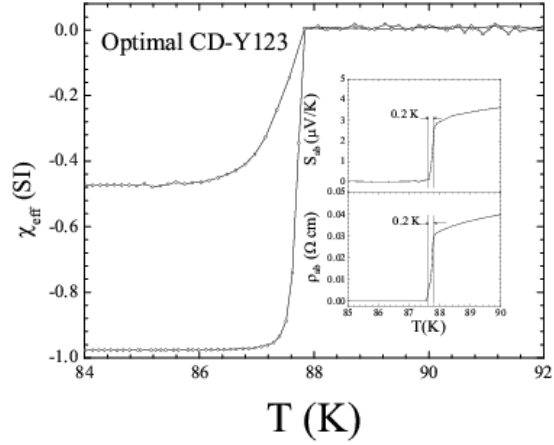

Figure 5. Zero-field cooled and field cooled susceptibilities vs. temperature for CD-Y123 at optimal doping level. Insets show superconducting transition of TEP and in-plane resistivity.

**4. Conclusions**

We have grown high quality calcium and oxygen co-doped $Y_{1-x}Ca_xBa_2Cu_3O_{7-\delta}$ single crystals using self-flux method. A quantitative *in situ* monitoring procedure has been established, which allows us to fine tuning the oxygen content and ensuring the preparation of equilibrated crystals. The crystals prepared following this procedure shows that the superconducting transition widths of TEP and resistivities are the smallest and the Meissner signal is the largest for CD-Y123 up to date. These results indicate we have achieved equilibrium crystals of high quality. Through studying hole density dependence of physical properties of CD-Y123 crystal will provide a alternative route for studying the intrinsic physical properties of high $T_c$ cuprates.

**Acknowledgments**

The authors acknowledge the kind assistance of J. K. Meen in WDS measurements and Y. Y. Sun in rocking curve measurements. This work was supported by the State of Texas through the Texas Center for Superconductivity at University of Houston.